\documentclass[a4paper,11pt]{article}

\usepackage{pos}
\usepackage{lipsum} 
\usepackage{subfig}
\usepackage{amsmath}
\usepackage{graphicx}
\usepackage{wrapfig}

\newsavebox{\measurebox}

\title{Performance Studies of the Acoustic Module for the IceCube Upgrade}

\ShortTitle{Performance Studies of the Acoustic Module for the IceCube Upgrade}

\author{The IceCube-Gen2 Collaboration \\{\normalsize \normalfont(a complete list of authors can be found at the end of the proceedings)}\\}

\emailAdd{cguenther@physik.rwth-aachen.de}

\abstract{

The IceCube Upgrade will augment the existing IceCube Neutrino Observatory by deploying 700 additional optical sensor modules and calibration devices within its center at a depth of 1.5 to 2.5\,km in the Antarctic ice. One goal of the Upgrade is to improve the positioning calibration of the optical sensors to increase the angular resolution for neutrino directional reconstruction. An acoustic calibration system will be deployed to explore the capability of achieving this using trilateration of propagation times of acoustic signals. Ten Acoustic Modules (AM) capable of sending and receiving acoustic signals with frequencies from 5 to 30\,kHz will be installed within the detector volume. Additionally, compact acoustic sensors inside 15 optical sensor modules will complement the acoustic calibration system. With this system, we aim for an accuracy of a few tens of cm to localize the Acoustic Modules and sensors. Due to the longer attenuation length of sound compared to light within the ice, acoustic position calibration is especially interesting for the upcoming IceCube-Gen2 detector, which will have a string spacing of around 240\,m. In this contribution we present an overview of the technical design of the Acoustic Module as well as results of performance tests with a first complete prototype.

\vspace{4mm}
{\bfseries Corresponding authors:}
Charlotte Benning$^{1}$, Jürgen Borowka$^{1}$, Christoph Günther$^{1*}$, Oliver Gries$^{1}$, Simon Zierke$^{1}$\\
{$^{1}$ \itshape RWTH Aachen University}\\[4mm]
$^*$ Presenter

\ConferenceLogo{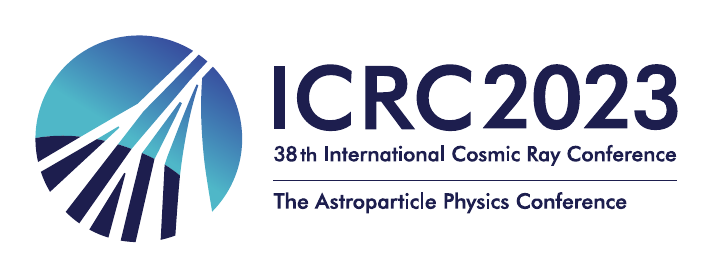}

\FullConference{The 38th International Cosmic Ray Conference (ICRC2023)\\ 26 July -- 3 August, 2023\\ Nagoya, Japan}
}

\begin{document}

\maketitle

\section{Introduction}\label{secInt}
\noindent
The IceCube Upgrade will improve the existing IceCube neutrino telescope by deploying approximately 700 new modules including photo sensors and multiple calibration devices at the bottom center of the detector. \cite{Ishihara:2019uL} These will be mounted on 7 new cable strings with a horizontal spacing of approximately 30\,m and about 2\,m vertically between modules along the strings.

The acoustic module (AM) is one of the new calibration devices. Ten AMs will be distributed over the Upgrade strings with distances ranging from a few tens of meters to 1000\,m. The main goal is to explore the feasibility of geometrical calibration of the photo sensors by measuring the acoustic propagation times between the modules in the ice. An accuracy in the order of 1\,ns $\times\,c_{\text{light,ice}} \simeq 20\,$cm is aimed as this will significantly improve the reconstruction of track-like signatures induced by high-energetic neutrinos \cite{Lilly2019}.

A good knowledge of the acoustic ice properties, including the speed of sound and attenuation, is important. The South Pole Acoustic Test Setup (SPATS) has already performed measurements of the speed of sound \cite{Abbasi_2010} and attenuation length \cite{Abbasi_2011} down to a depth of $\approx 500\,$m. The AM will enhance these results by measuring down to a depth of 2.5\,km with improved accuracy.

The AMs operate as high-power sound emitters and receivers operating at frequencies in the nominal range of 5-30\,kHz. The modules are integrated into the IceCube infrastructure and data-acquisition (DAQ). This allows timing accuracy of a few microseconds or better between the modules. The acoustic signals emitted by one module can be received by the others and propagation times are extracted from the acoustic waveforms using a dedicated analysis that extracts the group delay of chirp signals \cite{borowka2021acoustic}. From the propagation times, the positions of the AMs are reconstructed using a likelihood fit. Acoustic sensors placed in some of the optical modules, namely the pDOMs \cite{DuVernois:201613}, will allow the cross-calibration with optical methods and eventually a re-calibration of the existing IceCube detector. More details on these sensors can be found in \cite{Shefali:2019rR}.

Using acoustics for geometry calibration is promising as measurement results from SPATS indicate a long attenuation length in ice of up to a few hundred meters \cite{Abbasi_2011}. This is especially interesting for the IceCube Gen-2 detector with a planned string spacing of $\approx 240\,$m \cite{Aartsen_2021}. Also, acoustics can be operated simultaneously to optical operation, allowing a large repetition rate to improve the signal-to-noise ratio (SNR) of the recorded waveforms.

In this paper, we present measurements with the first prototype of the AM which have been carried out in a local swimming pool. The emitting power and receiving sensitivity as function of frequency calculated from these measurements are used to estimate the nominal distance range for operation in the Antarctic ice.

\section{Technical Design}\label{secTech}

\begin{figure}[ht]
    \centering
    \subfloat{{\includegraphics[width=7 cm]{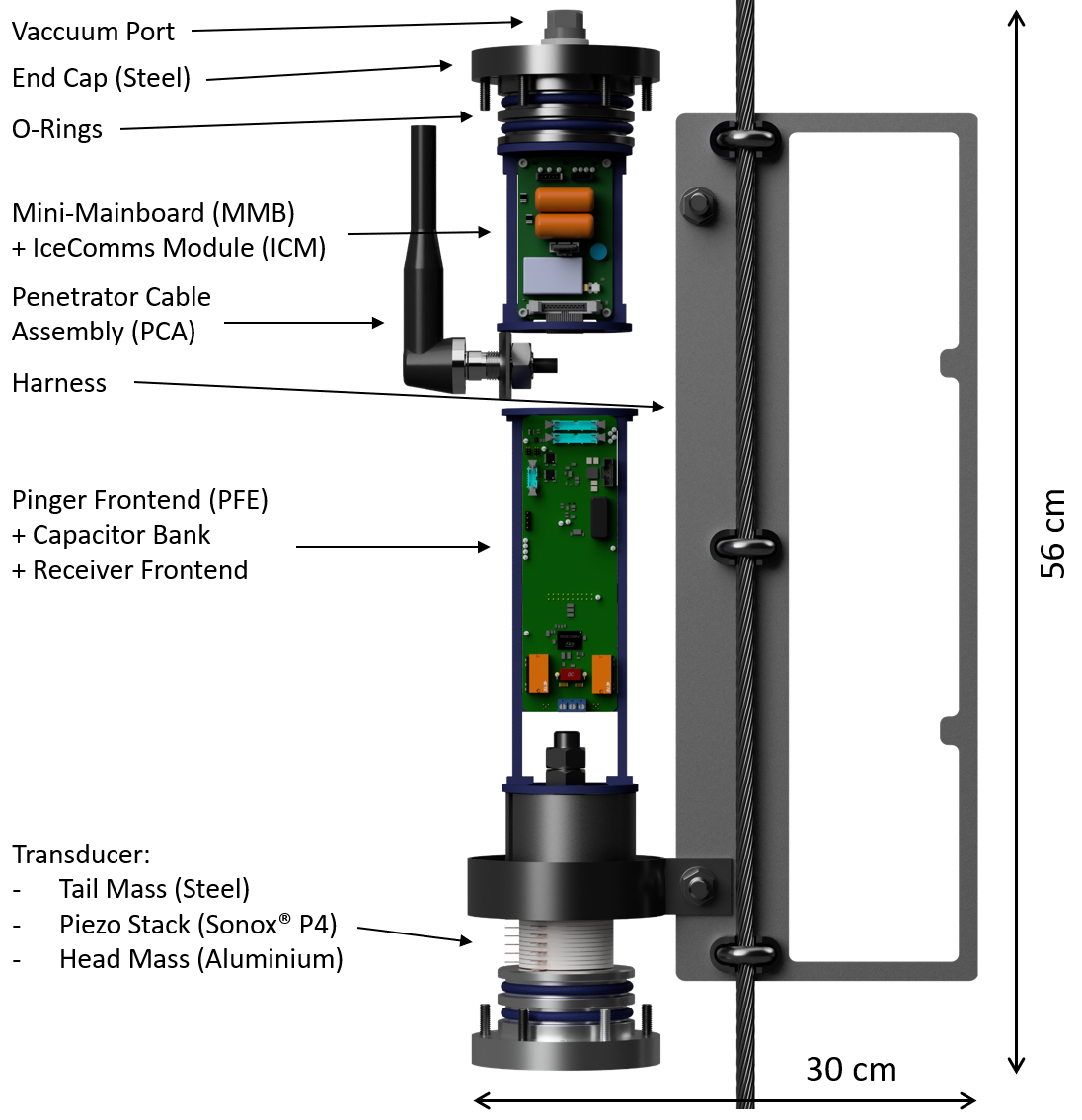}}\label{fig:AM}}
    \qquad
    \centering
    \subfloat{{\includegraphics[width=7.3 cm]{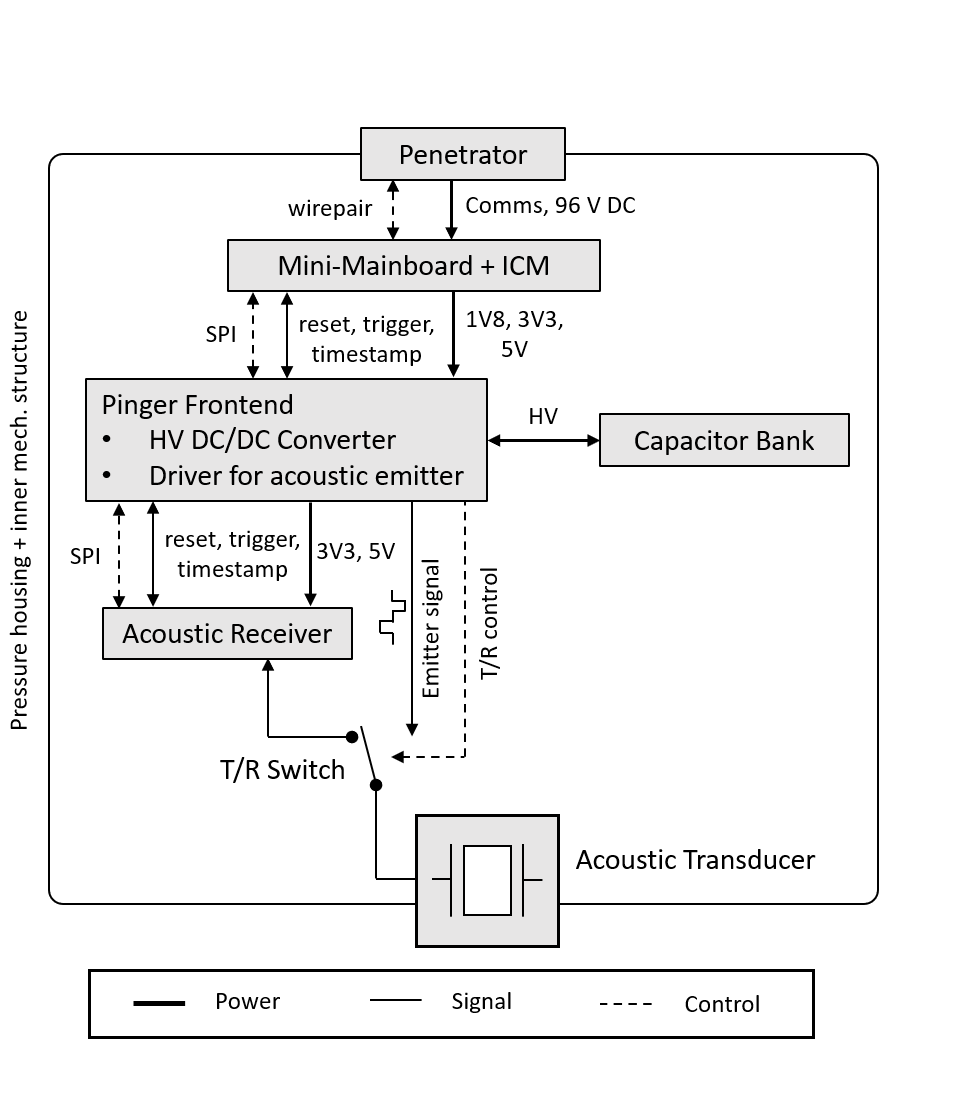}}\label{fig:diagram}}
    \caption{\textbf{(a)} Internal view of the acoustic module. The cylindric steel hull and the upper harness ring are omitted for clarity. \textbf{(b)} Block diagram showing the internal and external connections of the AM.}
    \label{fig:hardware}%
\end{figure}

The components of the AM are housed inside a powder-coated steel housing with a wall thickness of 1.5\,cm, ensuring pressure resistance of up to 700\,bar. A vacuum port allows to compensate for the atmospheric pressure difference between the production site and Antarctica to assure an under pressure inside the housing at all times. Cables for power supply and communication are fed into the module by a so-called Penetrator Cable Assembly (PCA), which is standardized for all IceCube Upgrade devices. All interfaces to the housing are sealed by multiple high-performance O-rings. Figure \ref{fig:AM} shows an illustration of the AM and its internal components.

The electronic components consist of the Mini-Mainboard (MMB), Ice Comms Module (ICM), Pinger Front-end (PFE), capacitor bank, and receiver front-end. The MMB is used in multiple devices in the Upgrade and is responsible for command and data handling. It consists of two boards, the controller and the power board. The power board is the interface to the main in-ice cable and supplies all other components in the AM with power. The controller board has two main components: an STM32H7 microcontroller, which controls the front-end boards, and the ICM, which handles the communication to the surface, the timing as well as the power distribution.

The PFE generates the emitter signals to drive the acoustic transducer. It charges a ceramic capacitor bank ($320\,\mu$F) using a high voltage (HV) DC/DC converter with up to 320\,V in about $10\,$s. A full-bridge driver uses the energy stored in the capacitor bank to generate bipolar rectangular signals with frequencies from 5-30\,kHz at a sampling rate of 1\,MS/s. A sine wave is approximated by the 4 output states of the full-bridge driver (+HV, 0, -HV, 0). Therefore, possible signal frequencies are $f_i = 1\,\text{MHz}/(4\cdot i), i>0$.

Relays on the PFE board allow switching between emitter and receiver mode. In receiver mode, only the receiver front-end is connected to the transducer and acoustic signals can be recorded with a sampling rate of up to $140\,$kS/s. The gain of the receiver is $\approx 53$\,dB @ 10\,kHz and has a bandwidth of 5-30\,kHz (@-3\,dB). The gain can be adjusted by software to adapt to the level of the acoustic signals. Figure \ref{fig:diagram} shows a block diagram of the interconnections within the module.

The acoustic transducer is a Tonpilz-style piezo transducer. It consists of a stack of 16 piezo discs clamped between an aluminum head-mass ($\approx0.635$\,kg), which also acts as the enclosure cap of one side, and a steel tail-mass ($\approx1.635$\,kg). The resulting mass ratio of $M_{\text{head}} : M_{\text{tail}} \approx 1:2.6$ increases the amount of sound emitted outwards of the module \cite{Lars2019}. The transducer has a resonance frequency at $\approx 10\,$kHz.

The AM will be attached to the Upgrade strings by a custom harness. It consists of two rings holding the AM which are connected to a mounting plate. Using three wire rope clamps the plate is attached to a 2\,m steel rope with thimbles. Only the lowest clamp is fully tightened to prevent stress on the steel rope due to thermal expansion. Stoppers above and beyond the clamps prevent the module from slipping along the rope. A bracket allows to fixate the main cable. All of the harness components are made from stainless steel. More details on the technical design of the acoustic module can be found in \cite{borowka2021acoustic}.

\section{Output Power\label{secEmit}}

The acoustic output power is measured by the Transmitting Voltage Response (TVR), which is defined as the Sound Intensity Level (SIL) generated at a distance of 1\,m by the transducer per 1\,V of input voltage in dependence of frequency \cite{butler2016transducers}:
\begin{equation}
    \text{TVR $(f)$ [dB\,re\,$\mu$Pa\,/\,1\,V\,@\,1\,m]} = \text{SIL}(f)\,\text{[dB\,re\,$\mu$Pa\,@\,1\,m]} - V_{\text{in}}\,\text{[dB\,re\,1\,V]}.
    \label{eq:TVR}
\end{equation}

The SIL can be measured by the output voltage of a transducer with known Open Circuit Receiving Response (OCRR). These are related by:

\begin{equation}
    \text{SIL $(f)$ [dB\,re\,$\mu$Pa]} = V_{\text{out}}\,\text{[dB\,re\,1\,V]} - \text{OCRR $(f)$ [dB\,re\,1\,V\,/\,$\mu$Pa]}.
    \label{eq:SIL}
\end{equation}

To measure the TVR, the AM is placed together with two absolutely calibrated ITC1001 hydrophones \cite{ITC1001} in a large water volume (swimming pool). The output power of the AM is compared to that of an Autonomous Pinger Unit (APU), which has been developed within the EnEx-RANGE project for acoustic trilateration in glacial ice. The transducer design of the APU is very similar to that of the AM. Differences are the number of piezo discs (8 compared to 16) and the head-to-tail mass ratio (3:1 compared to 1:2.6) \cite{EnEx_2020}.

\begin{figure}[ht]
    \centering
    \includegraphics[width=13 cm]{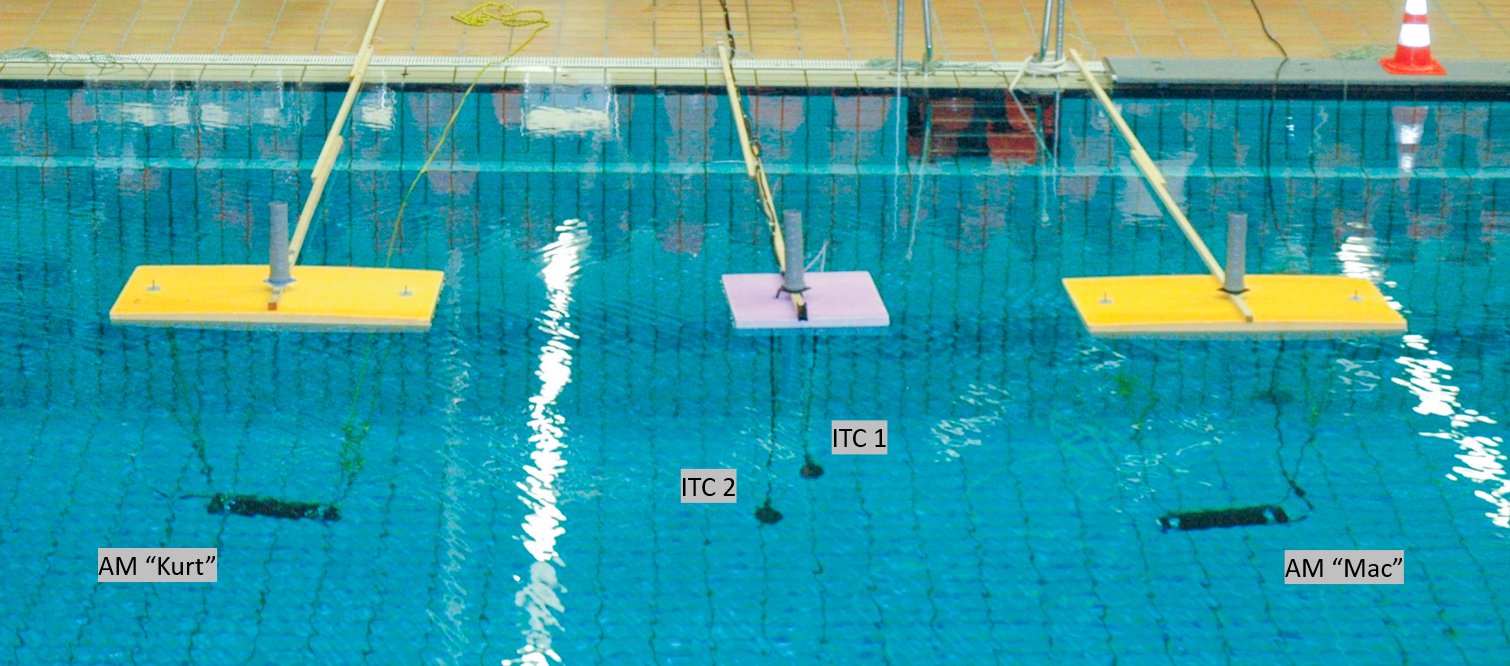}
    \caption{Picture of the measurement setup in the swimming pool. The AMs are beneath the two orange floatation panels and the two hydrophones are below the purple panel in the middle.}
    \label{fig:SBT}
\end{figure}

Figure \ref{fig:SBT} shows a picture of the measurement setup. The ITCs are placed in the center and separated by a distance of $\approx 1.5\,$m from each module. The distance of the closest wall of the swimming pool is $\approx 2.5\,$m. The exact locations of the probes have been measured to a precision of 10\,cm using a laser odometer and reference tubes on top of the floating panels. For the emitter measurements, the right AM ("Mac") is emitting and the two ITCs are receiving the acoustic signals. The setup for the APU measurements is analogous.

During the emitter measurement, the AM and APU are charged to a voltage of $300\pm 10$\,V and emit sine bursts with a duration of 10\,ms (5\,ms for the APU). The frequency of the bursts is varied between 5 and 30\,kHz. The ITCs are connected directly to an oscilloscope which triggers at a threshold of 100\,mV on the rising edge of the output signal. Typical recorded amplitudes range from a few 100\,mV to 2\,V. Five waveforms are recorded for each frequency and are averaged in later analysis.

\begin{figure}[ht]
    \centering
    \includegraphics[width=15 cm]{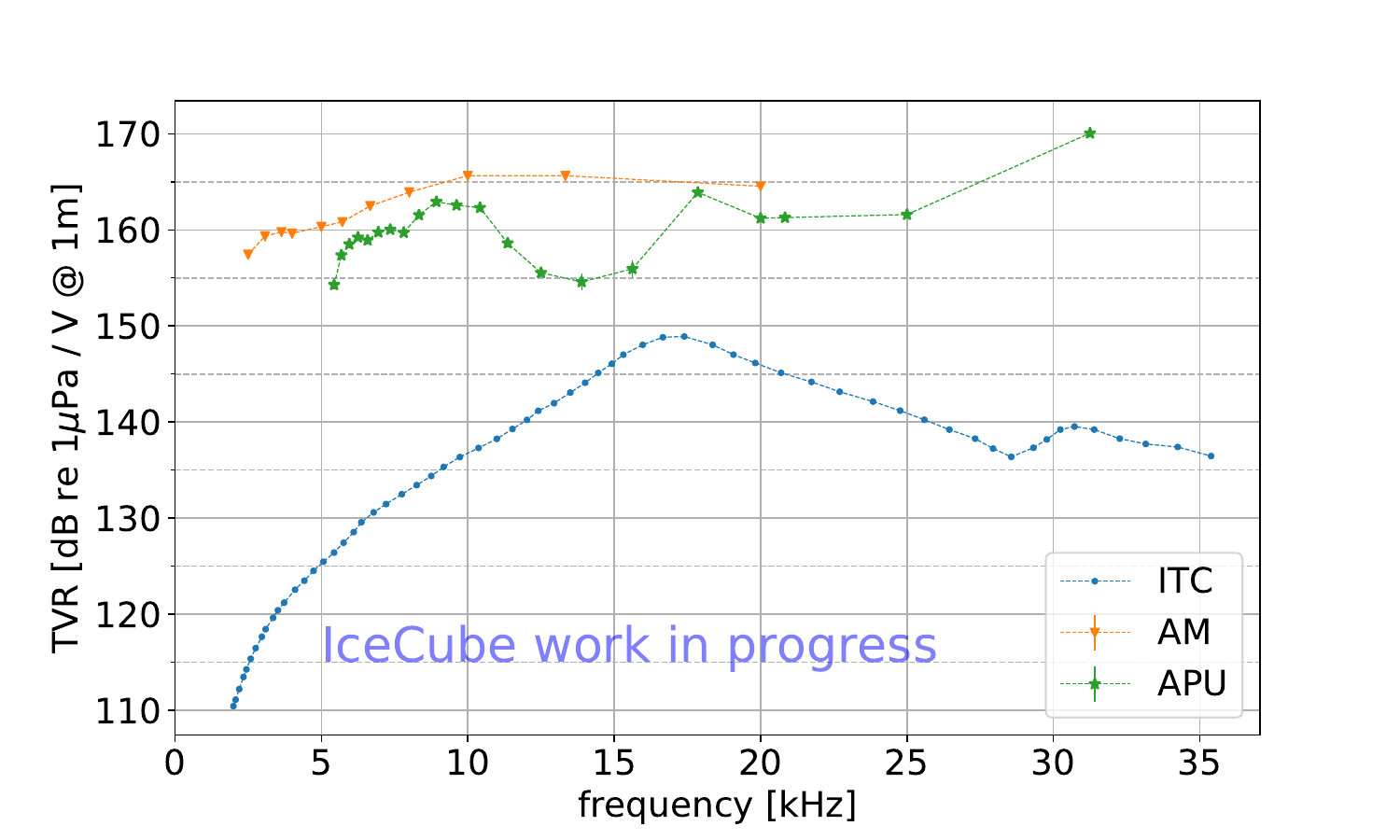}
    \caption{TVR for the acoustic module, APU and ITC-1001. Data for the ITC are from the datasheet \cite{ITC1001}.}
    \label{fig:TVR}
\end{figure}

The SIL at the ITC is measured by the voltage root-mean-square (RMS) of the waveforms. Therefore, the baseline is subtracted and the RMS is computed in a window of 1\,ms after the trigger threshold. The TVR of the AM and APU are calculated using equation \ref{eq:TVR}. The TVR has been scaled to a distance of 1\,m from pinger to ITC by assuming an attenuation of $1/d$: $\text{TVR @1\,m} = \text{TVR @\,d} - 10\cdot \log_{10} \left( \frac{d}{1\,\text{m}} \right)$.

The resulting TVR curves for the AM and the APU as well as for the ITC are shown in figure \ref{fig:TVR}. The maximum output is reached for both of the modules at $\approx$ 10\,kHz, which is the expected resonance frequency of the transducer. The output power of the AM is approximately 2-3 times higher compared to the APU. This can be explained by the increased number of piezo discs and the improved mass ratio of the transducer.

\newpage
\section{Range Estimation}\label{secRange}


The range of the AM is determined by the distance at which a minimum required SNR can be reached. During a measurement at the Langenferner glacier in Italy, the APU achieved an SNR of 100 at a distance of $\approx 30$\,m from APU to APU by averaging 64 waveforms \cite{EnEx_2020}. The attenuation length of the glacial ice was measured to be $\approx 8.85\pm0.95\,$m \cite{tc-13-1381-2019}. In the Antarctic ice, however, measurements by SPATS at shallow depths ($\approx 500\,$m) resulted in an attenuation length of approx. 300\,m \cite{Abbasi_2011}. Knowing that the AMs are similar to the APU and that the ice is similar to SPATS, although a slightly smaller attenuation length is expected for the warmer ice at deeper depths, we expect an attenuation length somewhere in-between. Combining this information with the measurements described in the previous sections, an estimation of the AM's range is drawn.

\begin{figure}[ht]
    \centering
    \includegraphics[width=15 cm]{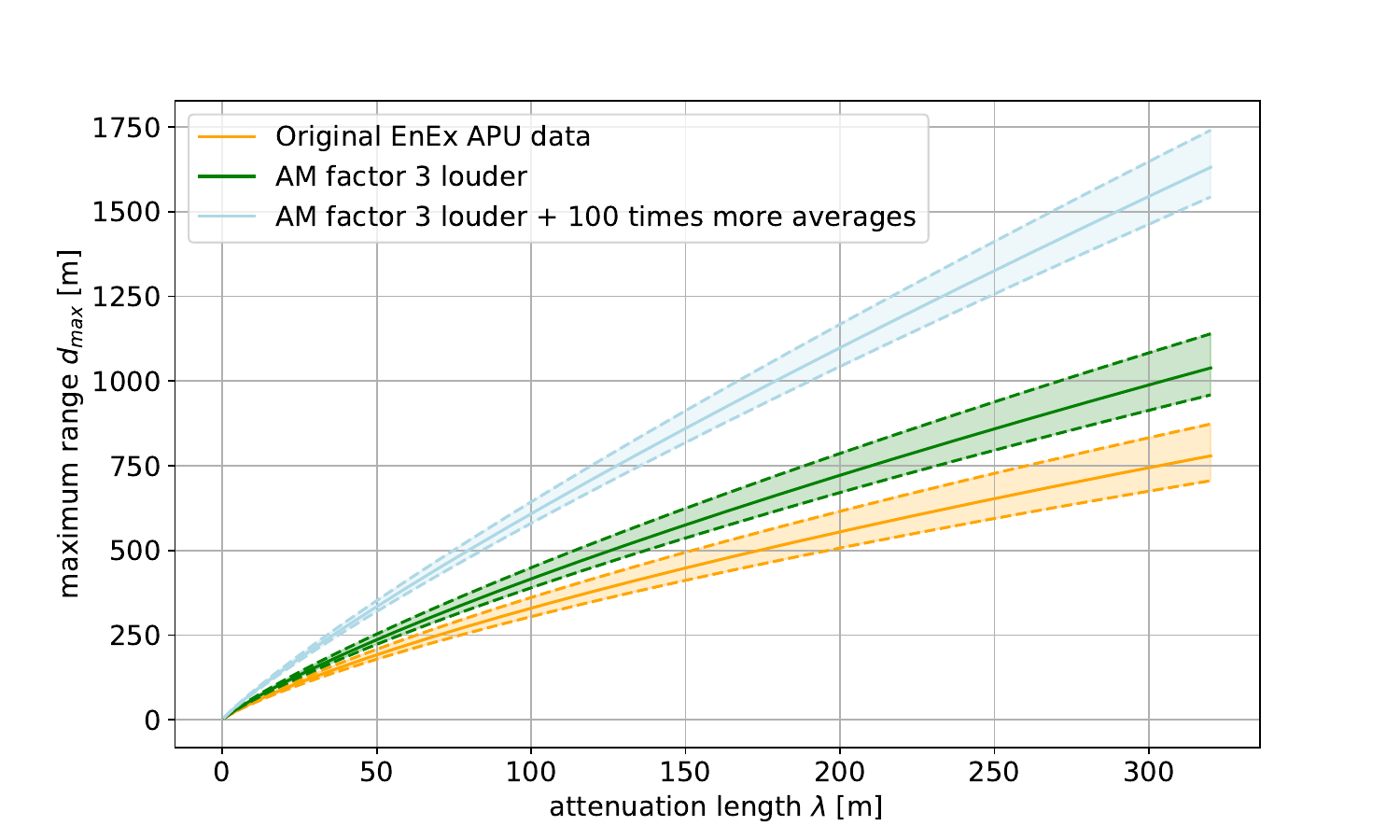}
    \caption{Expected range in dependence of the attenuation length of the AM and APU for the original measurement results by the EnEx measurements, assuming a factor of 3 more output power and for also assuming 100 times more averaging (6400 instead of 64).}
    \label{fig:range}
\end{figure}

The acoustic signal amplitude $A$ and therefore the SNR is attenuated by a geometrical factor $1/d$ and an exponential attenuation: $A(d) = \frac{a_0}{d}\cdot\exp{\left(-\frac{d}{\lambda}\right)}$, whereas $a_0$ can be fitted with the measured data. Solving this for $d$ and requiring a minimal SNR of 10:1 this equation can be solved for the maximal achievable range $d_{\text{max}}$:
\begin{equation}
    d_{\text{max}} = \lambda \cdot W\left(\frac{a_0}{A_{\text{min}}\cdot\lambda}\right),
\end{equation}
whereas $W(x)$ is the Lambert \textit{W} function and $A_{\text{min}}$ the minimum required signal amplitude. Figure \ref{fig:range} shows the resulting estimated range vs. attenuation length for the APU (orange) and also for assuming a factor of 3 more acoustic output power of the AM (green) as indicated by the measurement results. One can see that for an attenuation length of 300\,m as indicated by measurements of SPATS a range of $\approx 1000$\,m is expected. By increasing the number of averages, the SNR can be increased proportionally to $\sqrt{N}$. For a more conservative estimate of 175\,m for the attenuation length the same range is expected by averaging 100 times more waveforms. This is shown by the blue curve.

\section{Conclusion}\label{secConc}
The in-water measurements indicate an improved output power of the AM by a factor of 2-3 compared to the APU. This indicates an estimated range of the AMs in the Antarctic ice of up to 1000\,m or more. This is sufficient for the Upgrade configuration, which has a maximum distance between AMs of $\approx 1000\,$m. For the upcoming Gen-2 detector with a string spacing of 240\,m this estimate makes acoustic calibration an interesting option for geometric calibration. Results from the Upgrade will give insight into important acoustic parameters of the Antarctic ice at unreached depths and will explore the feasibility for future acoustic in-ice calibration systems.

\bibliographystyle{ICRC}
\bibliography{references}

\clearpage

\section*{Full Author List: IceCube-Gen2 Collaboration}

\scriptsize
\noindent
R. Abbasi$^{17}$,
M. Ackermann$^{76}$,
J. Adams$^{22}$,
S. K. Agarwalla$^{47,\: 77}$,
J. A. Aguilar$^{12}$,
M. Ahlers$^{26}$,
J.M. Alameddine$^{27}$,
N. M. Amin$^{53}$,
K. Andeen$^{50}$,
G. Anton$^{30}$
% ...
\\
\\
% ...
$^{12}$ Universit{\'e} Libre de Bruxelles, Science Faculty CP230, B-1050 Brussels, Belgium \\
% ...
$^{17}$ Department of Physics, Loyola University Chicago, Chicago, IL 60660, USA \\
% ...
$^{22}$ Dept. of Physics and Astronomy, University of Canterbury, Private Bag 4800, Christchurch, New Zealand \\
% ...
$^{26}$ Niels Bohr Institute, University of Copenhagen, DK-2100 Copenhagen, Denmark \\
$^{27}$ Dept. of Physics, TU Dortmund University, D-44221 Dortmund, Germany \\
% ...
$^{30}$ Erlangen Centre for Astroparticle Physics, Friedrich-Alexander-Universit{\"a}t Erlangen-N{\"u}rnberg, D-91058 Erlangen, Germany \\
% ...
$^{47}$ Dept. of Physics and Wisconsin IceCube Particle Astrophysics Center, University of Wisconsin{\textendash}Madison, Madison, WI 53706, USA \\
% ...
$^{50}$ Department of Physics, Marquette University, Milwaukee, WI, 53201, USA \\
% ...
$^{53}$ Bartol Research Institute and Dept. of Physics and Astronomy, University of Delaware, Newark, DE 19716, USA \\
% ...
$^{76}$ Deutsches Elektronen-Synchrotron DESY, Platanenallee 6, 15738 Zeuthen, Germany  \\
$^{77}$ Institute of Physics, Sachivalaya Marg, Sainik School Post, Bhubaneswar 751005, India \\
% ...

\subsection*{Acknowledgements}

\noindent
The authors gratefully acknowledge the support from the following agencies and institutions:
USA {\textendash} U.S. National Science Foundation-Office of Polar Programs,
U.S. National Science Foundation-Physics Division,
U.S. National Science Foundation-EPSCoR,
Wisconsin Alumni Research Foundation,
Center for High Throughput Computing (CHTC) at the University of Wisconsin{\textendash}Madison,
Open Science Grid (OSG),
Advanced Cyberinfrastructure Coordination Ecosystem: Services {\&} Support (ACCESS),
Frontera computing project at the Texas Advanced Computing Center,
U.S. Department of Energy-National Energy Research Scientific Computing Center,
Particle astrophysics research computing center at the University of Maryland,
Institute for Cyber-Enabled Research at Michigan State University,
and Astroparticle physics computational facility at Marquette University;
Belgium {\textendash} Funds for Scientific Research (FRS-FNRS and FWO),
FWO Odysseus and Big Science programmes,
and Belgian Federal Science Policy Office (Belspo);
Germany {\textendash} Bundesministerium f{\"u}r Bildung und Forschung (BMBF),
Deutsche Forschungsgemeinschaft (DFG),
Helmholtz Alliance for Astroparticle Physics (HAP),
Initiative and Networking Fund of the Helmholtz Association,
Deutsches Elektronen Synchrotron (DESY),
and High Performance Computing cluster of the RWTH Aachen;
Sweden {\textendash} Swedish Research Council,
Swedish Polar Research Secretariat,
Swedish National Infrastructure for Computing (SNIC),
and Knut and Alice Wallenberg Foundation;
European Union {\textendash} EGI Advanced Computing for research;
Australia {\textendash} Australian Research Council;
Canada {\textendash} Natural Sciences and Engineering Research Council of Canada,
Calcul Qu{\'e}bec, Compute Ontario, Canada Foundation for Innovation, WestGrid, and Compute Canada;
Denmark {\textendash} Villum Fonden, Carlsberg Foundation, and European Commission;
New Zealand {\textendash} Marsden Fund;
Japan {\textendash} Japan Society for Promotion of Science (JSPS)
and Institute for Global Prominent Research (IGPR) of Chiba University;
Korea {\textendash} National Research Foundation of Korea (NRF);
Switzerland {\textendash} Swiss National Science Foundation (SNSF);
United Kingdom {\textendash} Department of Physics, University of Oxford.

\end{document}